\documentclass[%
 aip,
 amsmath,amssymb,
 reprint,%
]{revtex4-1}

\usepackage{graphicx}
\usepackage{dcolumn}
\usepackage{bm}
\usepackage{tabularx}
\usepackage[utf8]{inputenc}
\usepackage[T1]{fontenc}
\usepackage{mathptmx}
\usepackage{color,soul}
\usepackage{float} 
\usepackage{url}

\begin{document}

\preprint{AIP/123-QED}

\title{Beyond optimization - supervised learning applications in relativistic laser-plasma experiments}
\author{Jinpu Lin}
 \email{linjinp@umich.edu.}
\author{Qian Qian}
\author{Jon Murphy}
 \affiliation{Center for Ultrafast Optical Science, University of Michigan, Ann Arbor, Michigan 48109, USA}
\author{Abigail Hsu}
 \affiliation{Department of Applied Mathematics and Statistics, Stoney Brook University, Stony Brook, NY 11794, USA}
\author{Yong Ma}
 \affiliation{Center for Ultrafast Optical Science, University of Michigan, Ann Arbor, Michigan 48109, USA}
\author{Alfred Hero}
 \affiliation{Department of Electrical Engineering and Computer Science, University of Michigan, Ann Arbor, Michigan 48109, USA}
\author{Alexander G.R. Thomas}
\author{Karl Krushelnick}
 \affiliation{Center for Ultrafast Optical Science, University of Michigan, Ann Arbor, Michigan 48109, USA}

\date{\today}

\begin{abstract}
We explore the applications of machine learning techniques in relativistic laser-plasma experiments beyond optimization purposes. We predict the beam charge of electrons produced in a laser wakefield accelerator given the laser wavefront change caused by a deformable mirror. Machine learning enables feature analysis beyond merely searching for an optimal beam charge, showing that specific aberrations in the laser wavefront are favored in generating higher beam charges. Supervised learning models allow characterizing the measured data quality as well as recognizing irreproducible data and potential outliers. We also include virtual measurement errors in the experimental data to examine the model robustness under these conditions. This work demonstrates how machine learning methods can benefit data analysis and physics interpretation in a highly nonlinear problem of relativistic laser-plasma interaction.
\end{abstract}

\maketitle

\section{\label{sec:level1}Introduction}
High-repetition-rate laser systems have been widely used with evolutionary algorithms to solve optimization problems in the field of relativistic laser-plasma interactions, including laser wakefield acceleration \cite{he2015coherent, dann2019laser, lin2019adaptive}, ion acceleration \cite{smith2020optimizing, noaman2018controlling}, x-ray production \cite{streeter2018temporal}, terahertz generation \cite{hah2017enhancement}, laser filamentation \cite{englesbe2016control, lefebvre2018phase}, and laser focus optimization \cite{nayuki2005production, lin2018focus}. However, evolutionary algorithms usually provide little information other than a local optimum, which can be hard to interpret. Instead, machine learning methods can generate predictive models that reveal more information in the dataset and help understand the physical processes that occur in an experiment. 

The broader discipline of plasma physics has adopted various machine learning methods in recent years. For instance, supervised learning regression algorithms have been applied to inertial confinement fusion (ICF) experiments with growing interests, such as Deep Jointly-Informed Neural Networks \cite{gaffney2019making, hsu2020analysis, humbird2019parameter, maris2019finding} and Gaussian Process regressor \cite{hatfield2019using}. Another popular machine learning technique called Random Forest has found success in magnetic confinement fusion experiments for both classification and regression problems \cite{rea2018disruption, piccione2020physics}. In space physics, Gaussian Processes are used to classify solar wind plasmas into categories \cite{camporeale2017classification},  and deep neural networks are used to predict solar flares from sunspot data \cite{jiao2020solar, chen2019identifying}. Beyond supervised learning, plasma physicists find interest in other powerful and increasingly popular machine learning methods, such as transfer learning \cite{humbird2019transfer} and reinforcement learning \cite{witman2019sim, kain2020sample}. The laser-plasma community is starting to embrace machine learning techniques as well. Artificial neural networks are employed to analyze features in high-order-harmonic spectra \cite{gonoskov2019employing} and laser-induced-breakdown spectra \cite{li2020laser}. Our work explores the capability of machine learning techniques in the field of laser-wakefield acceleration using all the supervised learning methods mentioned above.

Laser wakefield accelerators (LWFAs), first proposed by Tajima and Dawson \cite{tajima1979laser}, provide a possible alternative to conventional particle accelerators at a substantially lower size and cost. Taking advantage of the electric fields in laser-produced plasmas, LWFA can reach acceleration gradients of 100 GeV/m, which is three orders of magnitude greater than those produced in conventional accelerators. Extensive studies have been performed to understand LWFA mechanisms and to experimentally demonstrate energetic electron beams \cite{malka2002electron, mangles2004monoenergetic, geddes2004high, faure2004laser, esarey2009physics}, and the highest electron energy achieved so far is 7.8 GeV \cite{gonsalves2019petawatt}. While the highest energy facilities usually fire a few shots a day, there has been growing interests in having higher repetition rate lasers with slightly lower peak power and corresponding plasma targets \cite{roso2018high, prencipe2017targets, salehi2019high, salehi2017mev, feister2019development}. One of the rationales to increase the repetition rate is to compare LWFAs to conventional accelerators that operate at tens of Hz or above. In addition, it allows meeting the requirements for statistical methods to have better control over experiments \cite{he2015coherent, dann2019laser, lin2019adaptive, kim2017stable, liu2014adaptive, tsai2018control, shalloo2020automation}. 

In this work, we have predicted the electron beam charges in LWFA given the laser wavefront modification caused by a deformable mirror using four supervised learning regression methods: Random Forest, Neural Network, Deep Jointly-Informed Neural Network, and Gaussian Process. We benchmark these models and demonstrate applications beyond optimization. We show that generating higher beam charges favors specific wavefront aberrations, which is revealed by ranking the feature importance in form of the Zernike decomposition. We analyze the experimental data quality by evaluating the model performance on every measured data point. We also characterize the model robustness against a range of virtual measurement error bars assigned to the experimental data.

\section{Data and Methods}
\subsection{Data}
\subsubsection{Experimental}
The dataset is taken from the recent experiments performed in the Gerard Mourou Center for Ultrafast Optical Science at the University of Michigan. The experiments were performed using the Lambda Cubed laser system which provides 35 fs, 20 mJ laser pulses at 480 Hz. 
The laser wakefield accelerated electron beam data used here comes from four separate experiment days when running optimization algorithms. The laser wavefront change was induced by a deformable mirror and recorded as the 37 actuators' voltages on the mirror surface. The electron beam was captured via a scintillator screen that was imaged by an electron-multiplying CCD camera (Andor Luca-R, 14-bit). 
The optics which deliver the laser to the experimental chamber were unchanged throughout the duration of the four days, and the alignment procedures are a routine that is carried out the same way each day.


\subsubsection{Pre-processing}
The dataset is pre-processed before the regression modeling. The dataset contains the information of the optical wavefront change of the driving laser caused by a deformable mirror as well as the electron beam charge of the accelerated electrons. The wavefront changed by the deformable mirror has 37 dimensions in space and can be described mathematically by a polynomial, known as the Zernike polynomial \cite{born2013principles}, to reduce the dimension. In this case, the coefficients of the first 5 layers (15 terms) in the Zernike polynomial can accurately reproduce the wavefront. The 15-dimensional vectors consisting of the Zernike coefficients are used as the input to our supervised learning models, while the electron beam charges are the output, normalized to the range (0, 1]. In the context of machine learning, the input is called a feature and the output is called a label. We have 208 data samples in total. The dataset is split into two subsets: $80\%$ of the data points are used to train the models while $20\%$ are for testing. The feature matrix in the training set is sphered so that its rows have zero sample mean and unity sample variance. The feature matrix in the test set is updated accordingly with the same transformation.

\subsubsection{Correlation}
In statistics, the correlation coefficient measures the dependence between two variables and its value falls in the range of [-1,1]. The absolute value represents the strength of the dependence while the sign represents the direction. We calculate the correlation between each feature (Zernike coefficients) and the electron beam charge in the test dataset, as is illustrated in Table. \ref{correlationMatrix}. Among all the Zernike coefficients, while $z_{10}$ has the largest magnitude of correlation. The correlation matrix will be compared to the machine learning model predictions in Table. \ref{featureImportanceTest}, \ref{featureImportanceAll} in Section. \ref{section:result}.
\begin{table}[htbp]
  \centering
  \caption{Correlation between Zernike terms and beam charge.}
  \resizebox{\columnwidth}{!}{
    \begin{tabular}{|r|r|r|r|r|r|r|r|r|r|r|r|r|r|r|r|}
    \hline
    $z_0$ & $z_1$ & $z_2$ & $z_3$ & $z_4$ & $z_5$ & $z_6$ & $z_7$ \\
    \hline
    0.51 & -0.33 & -0.28 & 0.29 & 0.14 & 0.21 & -0.15 & 0.37\\
    \hline\hline
    $z_8$ & $z_9$ & $z_{10}$ & $z_{11}$ & $z_{12}$ & $z_{13}$ & $z_{14}$ & charge \\  
    \hline
    -0.42 & 0.20 & -0.55 & 0.35 & 0.47 & -0.42 & -0.019 & 1 \\
    \hline
    \end{tabular}%
    }
  \label{correlationMatrix}%
\end{table}%

\subsection{Machine learning methods}
Four supervised learning regression methods are used to predict the electron beam charges based on laser wavefront changes. Supervised learning is a branch of machine learning, which learns a function that maps an input (feature) to an output (label) based on example input-output pairs in a training sample\cite{russell2002artificial}. In each of the following supervised learning methods, the model is trained on the training dataset recursively until it can accurately predict the labels using the features. The model performance is then characterized by the test dataset. 

\subsubsection{Random Forest}
The Random forest (RF) regressor\cite{liaw2002classification} is a popular bagged algorithm for high-dimensional and nonlinear regression. It is based on the concept of a decision tree, which splits the dataset along some dimensions and recursively divides the space into regions with similar labels. Being the most popular bagging (Bootstrap Aggregation) ensemble algorithm, random forest samples, with replacement at uniform probability, the original dataset $D$ into $m$ datasets ($D_1, D_2, ..., D_m$) with the same size as $D$. For instance, if the original dataset contains three samples $D=[a,b,c]$, then $D_1$ could be $[a,c,c]$. For each dataset $D_j$ in the forest, we train a full decision tree by splitting the data to $k<n$ dimensions. Only k features are to be considered when looking for the best split. Since the trees become much more different as they select different features, we have to increase the number of trees and average over individual regressors. This bagging process helps reduce variance effectively. Denote $h_i$ as the regressor of the $i_{th}$ tree, then the bagged regressor is \cite{kilian}:
\begin{equation}
\label{rfEq}
h_i=\frac{1}{m}\sum_{i=1}^m h_i(x)
\end{equation}

In this study, we implement the algorithm using the \textit{Sklearn.ensemble.RandomForestRegressor} library in Scikit-learn \cite{pedregosa2011scikit}. The hyper-parameters to tune are the number of trees, the maximum depth in a tree, and the maximum number of features when splitting. 

\subsubsection{Deep Neural Network}


A Deep Neural Network (DNN) is a feed-forward artificial neural network with multiple hidden layers. The goal of a feed-forward network is to approximate some function $f$ \cite{goodfellow2016deep}. A typical one hidden layer can be mathematically described with weight $w$ and bias $w_0$ by Eq.~(\ref{eq:nn1}):
\begin{equation}
\label{eq:nn1}
y = f(w_0+w_1^Tx)
\end{equation}

According to the Universal Approximation Theorem, a feed-forward network with even one hidden layer 
can approximate any continuous function from one finite dimensional space to another under some conditions \cite{hornik1991approximation}. Although practically it may lead to an infeasibly large layer and fail to generalize, DNNs are powerful function approximators when learned properly. Compared to shallow models, DNNs usually can extract better features and learn more effectively. 

In this work, we build a fully-connected five-layer DNN using the \textit{Tensorflow.Keras} library \cite{chollet2015keras} based on Google’s deep learning software TensorFlow \cite{abadi2016tensorflow}. When constructing the network, we use the rectified linear unit (ReLU) function and the Sigmoid function as the activation functions for different layers.
The cost function is the mean squared error loss governed by the Adam optimizer \cite{kingma2014adam} to update the network weights. A $L_2$ norm regularization is added to the loss function to reduce overfitting. The main tuning parameters are the number of layers, the number of neurons in each layer, the epoch size, and the initialization of the weight matrix.

\subsubsection{Deep Jointly-Informed Neural Networks}
Deep Jointly-Informed Neural Networks (DJINN) is a machine learning algorithm that constructs and initializes the deep feedforward neural networks based on decision trees. It was developed by Humbird \textit{et al.}\cite{humbird2018deep} and it has shown success in training ICF datasets \cite{humbird2019parameter, gaffney2019making, hsu2020analysis} as well as standard regression datasets such as Boston housing prices, California housing prices, and diabetes disease progression \cite{humbird2018deep}. The algorithm starts by constructing a decision tree or an ensemble of trees, where the number of trees will be the number of networks in the later stage. It then maps the decision trees to deep neural networks by taking the decision paths as guidance for the network architecture and weight initialization. The networks are trained with backpropagation using TensorFlow \cite{abadi2016tensorflow}. In the network architecture, the activation function is the rectified linear unit and the cost function is governed by Adam optimizer \cite{kingma2014adam}. Without optimizing the architecture of the neural networks, DJINN displays comparable performance to optimized architectures at a significantly lower computational cost using their datasets.

The DJINN regression source code is accessible at the LLNL/DJINN github directory. The main tuning parameters for this study are the maximum depth of trees, and the number of trees (nets) in ensemble.

\subsubsection{Gaussian Process}
The Gaussian process (GP) is a non-parametric Bayesian algorithm for supervised learning problems \cite{rasmussen2003gaussian}. While most machine learning algorithms fit the dataset into a model function with weight parameters and use that function to make predictions, Bayesian methods avoid the intermediate step and make predictions directly from the dataset. This is achieved by integrating all possible weight functions in the universe \cite{kilian}: 

\begin{equation}
\label{gpEq}
P(y|x,D)=\int _{w} P(y|x,w)\cdot P(w|D)\; dw
\end{equation}
where P is the prediction, w is the weight matrix, and D is the dataset. In Gaussian process regression, we assume that the data can be fit by some model with weight function w and a Gaussian distributed noise $\epsilon$: $y=f(x)=w^{T}x+\epsilon$. Thus the term $P(y|x,w)$ in Eq. \ref{gpEq} is a Gaussian distribution. The second term on the right-hand-side can also be proved to be Gaussian using the Bayes' rule:

\begin{equation}
\label{BayesEq}
P(w|D)=\frac{P(D|w)\;P(w)}{P(D)}
\end{equation}

where $P(w)$ is the prior distribution, $P(w|D)$ is the posterior distribution after D is considered, and $P(D)$ is a normalization. By choosing a Gaussian prior, Bayes' rule leads to a Gaussian posterior $P(w|D)$. Marginalizing out w in Eq. \ref{gpEq}, we know $P(y|x,D)$ is also in a Gaussian distribution. 
The Gaussian process has a closed form posterior distribution that can be used to quantify uncertainy of the estimate (posterior mode) through a confidence interval.

Another advantage of Gaussian process regression is that we can specify prior information about the shape of the model by selecting certain kernel functions. As is introduced previously, the probability function of the prediction can be expressed by a Gaussian distribution: $P(y|x,D)=N(\mu, \Sigma)$, where $\mu$ is the mean and $\Sigma$ is the covariance matrix. The covariance matrix can be kernelized using selected kernel functions. In this project, we implement the algorithm using the $Sklearn.gaussian\_process$ library in Scikit-learn \cite{pedregosa2011scikit} with a combination of Matern kernel and Rational Quadratic kernel. The hyper-parameters to tune are the smoothness, the length-scale, and the scale mixture parameter in the kernels.

\section{Results}
\label{section:result}
\begin{figure}[ht]
\centering
\includegraphics[width=0.9\columnwidth]{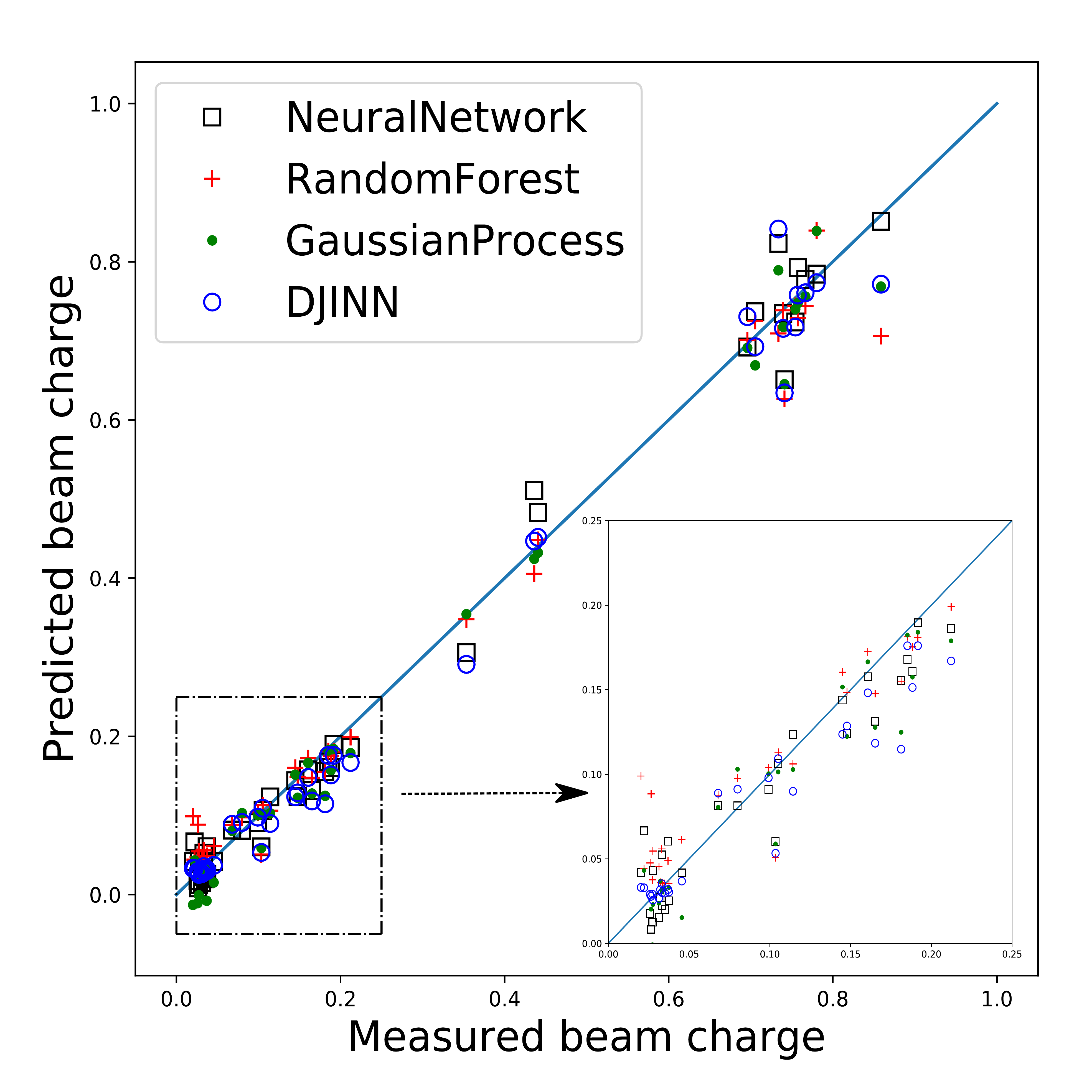}
\caption{Predicted vs. measured electron beam charge in test dataset.}
\label{predVSreal}
\end{figure}

All codes are written in Python. After training the models, we predict the electron beam charge using the laser wavefront change in the test dataset. Predicted electron beam charges using the above models are shown in Fig. \ref{predVSreal} against measured electron beam charges. A reference line at $45^{\circ}$ is included and data points closer to the reference line are considered better predictions. The bottom left corner of the plot is zoomed in and shown to the right. Detailed statistical evaluations are summarized in Table. \ref{all_results}, in which we report the mean-square-error (MSE), mean-absolute-error (MAE), R-squared ($R^2$) and explained variance score (ExVar) based on the predicted charge and the measured charge. MSE measures the average squared difference between the predictions and the real values, which contains information of both variance and bias. It is the most popular metric when evaluating machine learning models and we use MSE as the target for the hyperparameter tuning process. The downside of MSE is that it can be sensitive to outliers, which MAE handles better by measuring the absolute error instead of the squared error. $R^2$ is the proportion of variance of the measured value from the prediction. It tells how likely a new sample (out of the dataset) can be predicted by the model. Explained variance considers bias on top of $R^2$. It is the same as $R^2$ if the mean of error is 0. In general, we'd like to have MSE and MAE close to 0 while $R^2$ and ExVar close to unity. 

\begin{table}[ht]
\begin{tabular}{ |c|l|c|c|c| } 
 \hline
 Model &  MSE & MAE & $R^2$ & ExVar\\
\hline\hline
 Random Forest & 0.00132 & 0.0268 & 0.986 & 0.987 \\ 
\hline
  Neural network &0.00162 &0.0292  &0.983  &0.984 \\ 
 \hline
  DJINN & 0.00154 & 0.02741  & 0.98403  & 0.98404 \\ 
 \hline
  Gaussian Process & 0.00185 & 0.0305 & 0.981 & 0.981\\ 
 \hline
\end{tabular}
\caption{Evaluation matrix.}
\label{all_results}
\end{table}

All four models demonstrate similar statistics in Table. \ref{all_results}, though RF performs slightly better and GP gives the largest MSE and MAE scores while the smallest $R^2$ and ExVar scores. It also predicts negative values when the electron beam charges are small. However, it does not necessarily mean that Random Forest is the best model and gaussian Process is the worst. The results in the evaluation matrix are sensitive to the way we split the training set and test set. We will show in the next section that training and evaluating the model on different data points yields different results. We will also present more analyses such as the model consistency against measurement errors and overfitting-related issues.

\subsubsection{Data quality}

\begin{figure*}[ht]
\centering
\includegraphics[width=1.8\columnwidth]{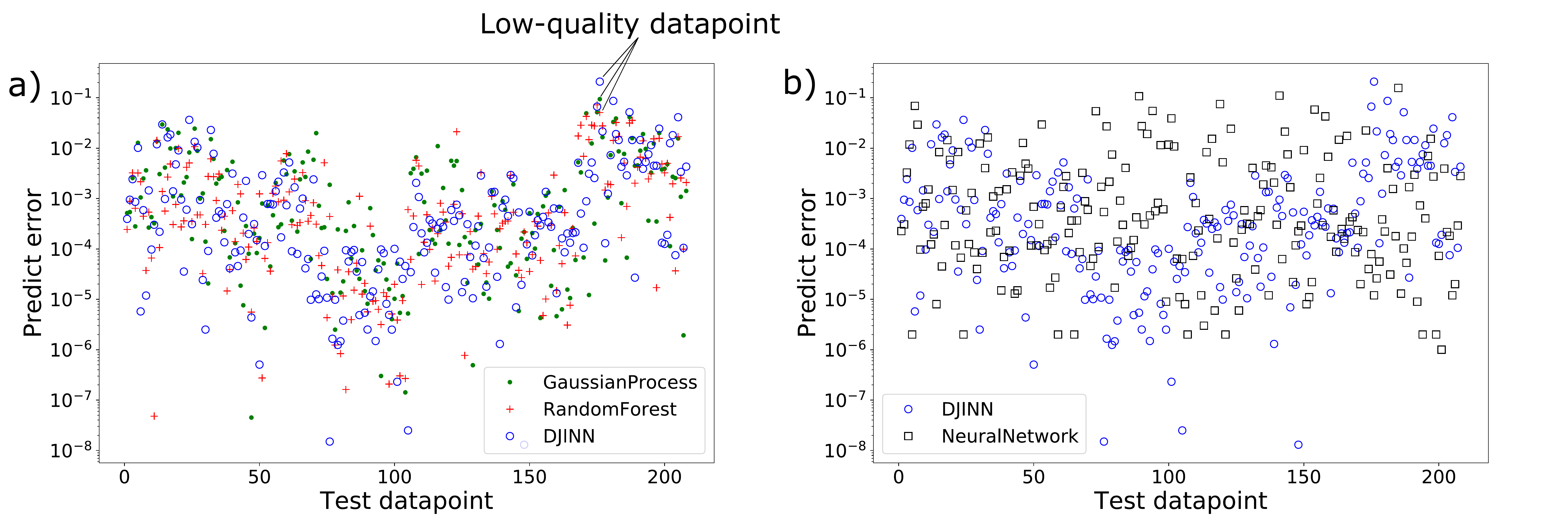}
\caption{Prediction error of every data point.}
\label{Dataquality}
\end{figure*}

Experimental measurements in LWFA can suffer from a lack of reproducibility and may have outliers in the dataset. Possible sources include the pointing instabilities and shot to shot fluctuations in high power laser systems, as well as the irreproducibility in the plasma density profile from gas jets. To justify the quality of our measured data, we make a prediction using the wavefront change of each data point and compare that to the corresponding measured electron beam charge. Instead of splitting the dataset into $80\%$ for training and $20\%$ for testing, here we test on only one data point while all the other data points are used to train the model. We then compare the predicted beam charge to the measured electron beam charge of this particular data point and calculate their difference. This process is looped over the entire dataset. Fig. \ref{Dataquality}a plots the prediction error ($\sigma=|y_{predicted}-y_{measured}|$) at each data point in random order using GP, RF, and DJINN. The prediction errors from DNN are shown in Fig. \ref{Dataquality}b. It is observed that the three models in Fig. \ref{Dataquality}a have similar performance while the results from DNN in Fig. \ref{Dataquality}b is very different. We interpret three messages from these two plots. 1. Prediction errors vary across seven orders of magnitude, i.e., some data points can be accurately predicted ($\sigma\sim1e-8$) while some data points can hardly be predicted ($\sigma\sim0.1$). Therefore selecting different data points into the training or test set can lead to different evaluation matrices from the one in Table. \ref{all_results}. 2. This huge variation can be caused by either the inconsistency of the model across data points, or this specific data being very different from all the other data points in the dataset that are used to train the model. We are satisfied with the reliability of the models, as is characterized in Table. \ref{all_results}. In addition, the similarity among the three models' performance in Fig. \ref{Dataquality}a suggests that the models are less likely to be inconsistent but in a similar manner. Thus it provides a potential characterization of the quality of each data point, i,e., a data point that has a very large prediction error in all three models can be considered as having poor data quality and may be dropped as an outlier. 3. DNN performs differently from the other models, suggesting that it overfits the data set and it is less reliable in this scenario. It matches the learning curves in Fig. \ref{learningcurve}, and a detailed discussion on the overfitting issue can be found later.

This could have some practical use in analyzing experimental results. For example, if we drive the same laser pulse into the wakefield accelerator twice, we might observe different electron beam charges due to a lack of reproducibility in LWFA. A typical solution would be to calculate the mean value and the error bar. However, this could include misinformation if there is a "bad" measurement that should be dropped. By performing the above analysis we would be able to tell which one of the measured beam charges is more reliable. Moreover, it can help identify outliers not only in repeated measurements but anywhere in the parameter space that the experiment scans across. The labeled data point in Fig. \ref{Dataquality}a is an example of possibly poor data quality, where all three models have large prediction errors at this point.

\subsubsection{Robustness against measurement errors}
\label{section:Virtual measurement error}
\begin{figure*}[ht]
\centering
\includegraphics[width=1.8\columnwidth]{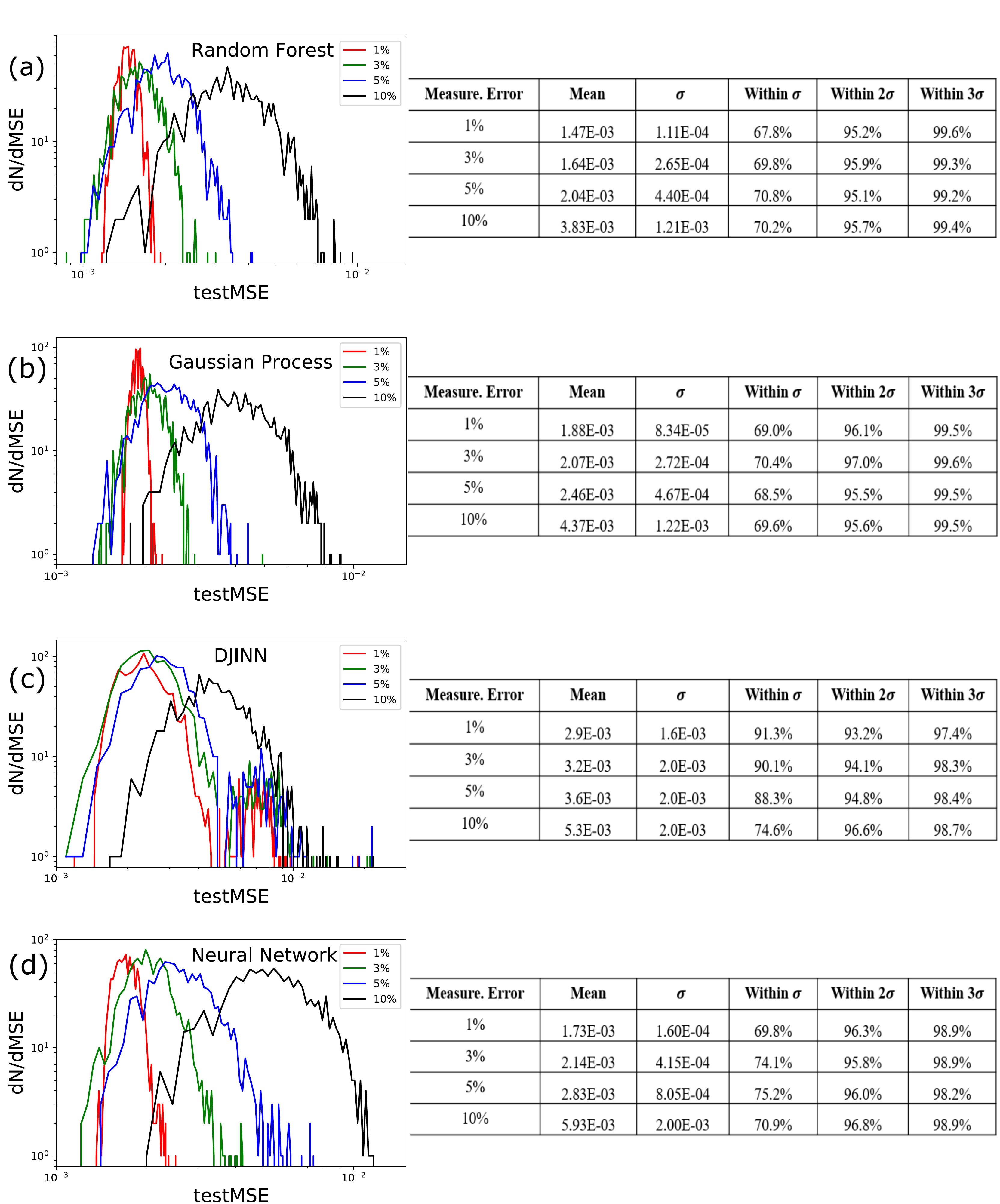}
\caption{Model performance against virtual measurement errors.}
\label{noisyData}
\end{figure*}

Studying science with error bars has been a rising trend as experimental measurements often come with such. In this section, we investigate the performance of these models against measurement errors of the electron beam charges. Since measurement errors were not recorded during our experiments, we include some virtual error bars to every measured electron beam charge. At each measurement, the true value is assumed to lie in the range of measured value $\cdot(1\pm X\%)$, where $X=1,3,5,10)$, and 1000 points are drawn randomly from a normal distribution within this range. Therefore we get 1000 copies of the original dataset with the same wavefront but different electron beam charges. The reason to have 1000 datasets is to generate enough statistics to justify the model performance against unsure measurements. Results are presented in Fig . \ref{noisyData}.

Fig. \ref{noisyData}a shows the distribution of the test MSE using RF. Each colored line is generated from 1000 MSEs. During the training process, the model configuration was kept the same among the 1000 datasets but the weight learning was updated in each dataset. Measurement errors that define the range of the dataset fluctuation are $1\%,\;3\%,\;5\%$ and $10\%$, while the corresponding test MSE distributions are plotted in red, green, blue and black, respectively. Statistical analysis is summarized in the adjacent table to the right, illustrating the mean value and standard deviation ($\sigma$) of the test MSE distribution as well as the percentage of points that fall within one, two, or three standard deviations around the mean value. Fig. \ref{noisyData}b-d present the results using GP, DJINN, and DNN. The four models share some common performances. The mean test MSE value increases with the measurement error, which means it is more likely to make a less accurate prediction when the measurement itself is less accurate, as expected. The standard deviation also increases with the measurement error, suggesting a less consistent or less precise model prediction at larger measurement errors. There are noticeable differences in the last three table columns of the four models. Remember that the virtual measurement errors are drawn from a perfect normal distribution, where the percentage of values that lie within one, two, or three standard deviations around the mean value are $68.3\%$, $95.5\%$, and $99.7\%$, respectively. As is shown in the tables, RF and GP retain almost normal distributions in the sample prediction, DNN gives normal-like distributions and the results from DJINN are far from normal distributions. 

\subsubsection{Learning curve}
\begin{figure}[ht]
\centering
\includegraphics[width=0.9\columnwidth]{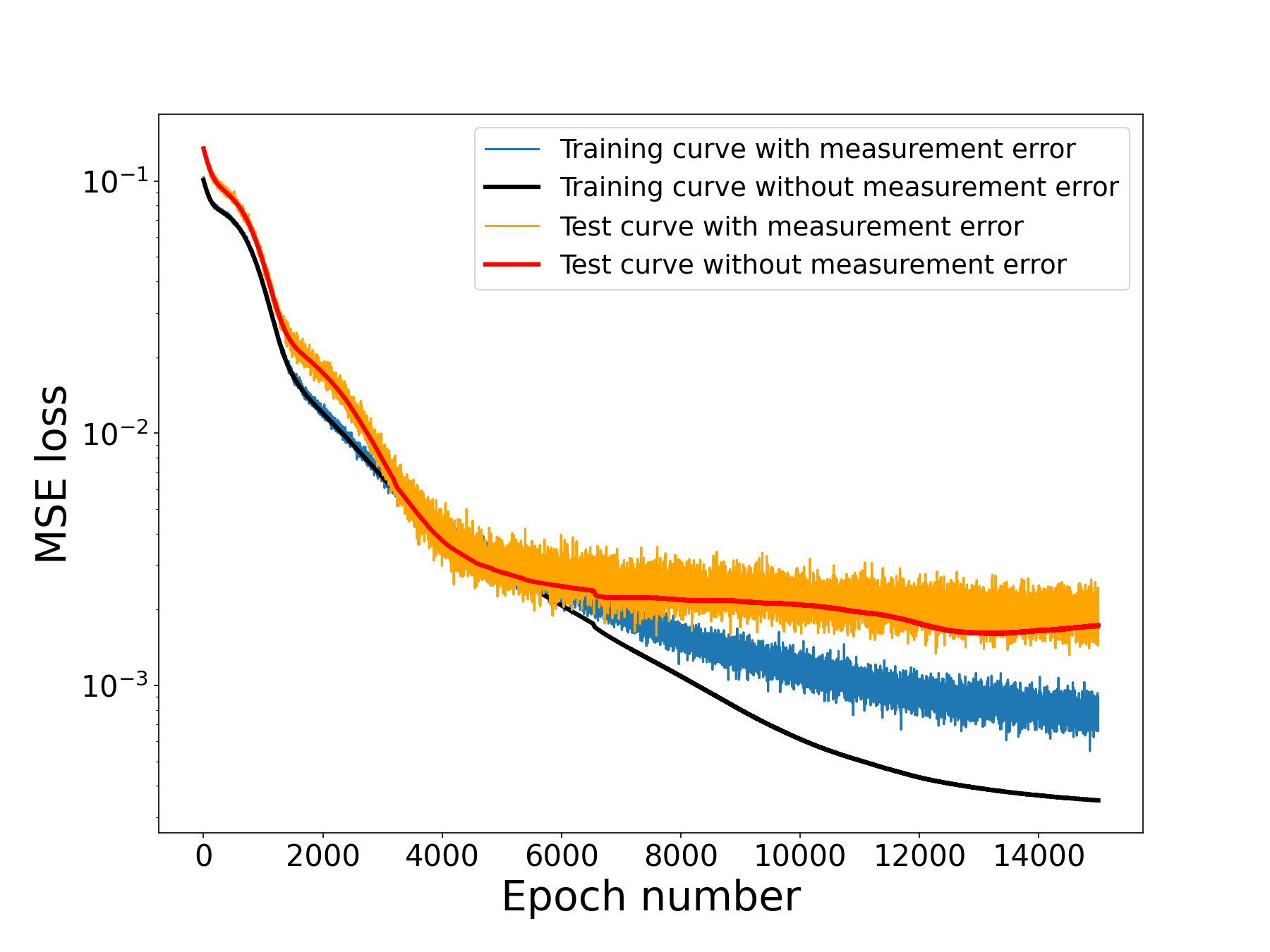}
\vspace*{-3mm}
\caption{Learning curves in Neural Network}
\label{learningcurve}
\end{figure}

In machine learning, overfitting often occurs when a model has learned the training data so well that it also learns the statistical noise or random fluctuations in the data. The learning curve is an intuitive tool to visualize the degree of overfitting. Fig. \ref{learningcurve} shows the learning curves in our Neural Network model, which plots MSE at each epoch for both training data (black) and test data (red). The training curve tells how well the model learns while the test curve tells how well the model generalizes. Since the red curve does not decrease as much as the black curve does, the model overfits. In other words, the overfitted model performs worse outside the training dataset. Fig. \ref{learningcurve} also plots learning curves considering measurement errors in blue and orange to better reproduce experimental conditions. Measurement errors are included in a similar way to the ones in the previous section, where the true value is assumed to lie in the range of measured value$\cdot(1\pm 3\%)$. Instead of generating 1000 copies of datasets at a time, here we generate only one dataset with measurement errors and update the measurement errors at every epoch in the learning process. Obtained learning curves are noisier but demonstrate less overfitting, as is shown in Fig. \ref{learningcurve}. After applying measurement errors, the training curve moves higher but the test curve does not shift much. Namely, the model finds it harder to learn but it is still able to make equally accurate predictions. Therefore, including virtual measurement errors is beneficial as it not only represents practical experimental conditions better but also decreases overfitting. Note that we have also tried this using the DJINN model but the learning curves with and without measurement errors look almost the same. It is not surprising since the DJINN model does not overfit as much as the neural network model does.      

\subsubsection{Feature importance}
We have trained models to predict the electron beam charge upon laser wavefront modification represented by the first 15 Zernike coefficients. It is natural to ask how sensitive the beam charge is to these features. We evaluate the feature importance using our four models and compare them to the correlation ranking, summarized in Table. \ref{featureImportanceTest} and \ref{featureImportanceAll}. In each row, we list the four most important features decided by that model, while the numbers are the orders of the Zernike coefficients. The importance of a feature is measured by calculating the increase in the model's prediction error after setting the feature values to a constant. A feature is said to be important if the prediction error increases significantly, and less important when the prediction error does not change much. When evaluating each feature, the values of this feature of all test data are set to their mean value. Note that model training is performed prior to this process and the training data are not modified. It is debatable whether the importance should be computed on the test data or the training data. The former tells how much the model relies on each feature for making predictions while the latter tells how much the feature contributes to the performance of the model on unseen data \cite{molnar2020interpretable}. We show the feature importance computed from both ways. In Table. \ref{featureImportanceTest} we split the dataset to a training set ($80\%$) and a test set ($20\%$) and evaluate the feature importance on the test set. In Table. \ref{featureImportanceAll} we train the model using the entire dataset and measure the feature importance also on the entire dataset.

Depending on the model and the evaluation data, the feature importance rankings in Table. \ref{featureImportanceTest} and \ref{featureImportanceAll} are slightly different. However, the $0^{th}$, $1^{st}$, $6^{th}$ and $10^{th}$ Zernike terms are generally believed to be the more important ones, physically representing the piston, tilt, vertical trefoil and oblique quadrafoil in wavefront aberration, respectively. It suggests that controlling these features would be more effective at producing high electron beam charges in this scenario. 


\begin{table}[ht]
\begin{tabular}{ |c|c|c|c|c| } 
 \hline
 Model & $1^{st}$ & $2^{nd}$  & $3^{rd}$ & $4^{th}$\\
\hline\hline
 Random Forest & 0 & 6  & 1 & 10 \\ 
\hline
 Gaussian Process &  1 &  10 & 0 & 6 \\ 
\hline
  Neural Network & 0 & 10 & 1 & 3 \\ 
 \hline
  DJINN & 1 & 10  & 0 & 6 \\
 \hline
  Correlation & 0 & 10 & 1 & 13 \\
 \hline
\end{tabular}
\caption{Evaluate feature (Zernike coefficient) importance on the test dataset.}
\label{featureImportanceTest}
\end{table}

\begin{table}[ht]
\begin{tabular}{ |c|c|c|c|c| } 
 \hline
 Model & $1^{st}$ & $2^{nd}$  & $3^{rd}$ & $4^{th}$\\
\hline\hline
 Random Forest & 0 & 6 & 1 & 12 \\ 
\hline
 Gaussian Process &  10 & 1 & 0 & 6 \\ 
\hline
  Neural Network & 0 & 10  &  1 & 8\\ 
 \hline
  DJINN & 1 & 10 & 0 & 6 \\
 \hline
 Correlation & 10 & 0 & 12 & 13 \\
 \hline
\end{tabular}
\caption{Evaluate feature (Zernike coefficient) importance on the entire dataset. }
\label{featureImportanceAll}
\end{table}

\section{Discussion}

There has been growing interest in the laser-plasma community to discuss what machine learning can bring us and what it can not. Machine learning methods are not expected to offer some generalized predictive models that save experimentalists from carrying out every experiment. This is due to the lack of reproducibility in high-power laser-plasma experiments, where laser systems usually suffer from shot-to-shot fluctuations and a given plasma density profile is hard to duplicate. However, using machine learning techniques can help us better understand the experiment we did and improve the design of next-step experiments. It enables deeper physics interpretation of the data since the predictive accuracy of the regression models is determined by the data quality. For example, by ranking the feature importance we are able to identify the Zernike terms that are most sensitive to noise. Some of them turned out to be the high order terms (vertical trefoil and oblique quadrafoil) that locate at the edge of the wavefront. This can be explained from an experimental point of view as follows. Deformable mirrors are manufactured to have actuators forming into a hexagon matrix, while a wavefront is usually defined in a circular or rectangular shape, leading to lack of information on the edge. If the wavefront is measured directly with wavefront sensors, it is usually required to manually draw a circle that covers most lighted pixels on the detector as the region of interest. As a result, uncertainty arises at the pixels on the edges of the region of interest. If the wavefront is reconstructed using the actuator displacement on the deformable mirror surface, the phase in these unknown edges also needs to be defined and is set to be nan in our case. Another possible source of noise is the imperfect overlap of the laser and the deformable mirror surface: either the laser beam clipped off the mirror edge or it did not fulfill the whole mirror surface.

Including virtual measurement uncertainties can be useful even if the experimental data come with some uncertainties. It may be able to narrow down the range of uncertainty when the measurement uncertainty is large. For instance, if a data point in the dataset has some large uncertainty and we would like to know or at least narrow down the true measurement value, machine learning regression methods can provide us a possible solution. We can start by randomly sampling N points within the range of uncertainty of this measurement value. The next step is to make N copies of the measurement dataset and replace the point of large uncertainty with one of our sampled point in each dataset. Therefore, we obtain N similar datasets with difference only at one point. We then train and test the machine learning models on each dataset. Those datasets who lead to less accurate predictions (large test MSE) are less likely to contain true measurement at the uncertain point if the models make sense.

It is also worth discussing the kernel functions in the Gaussian Process method. The common way of kernel selecting is either to have expert knowledge about the dataset or to compare candidate kernels for the best performance. In this project, we have tried different kernels and decide that a combination of Matern kernel and Rational Quadratic kernel works best. We can thus infer some knowledge about our dataset based on the rationale of these kernels. Matern is a generalization of the popular Gaussian radial basis function (RBF) with tunable smoothness \cite{rasmussen2003gaussian}. The smoothness of the model can be controlled by a parameter $\mu$ while the $\mu$ value in our case is as low as 0.3, suggesting that the resulting function is far from smooth. The other kernel that fits our model, the Rational Quadratic kernel, is a sum of RBF kernels with different length scales. Note that the length scale decides the safe distance to extrapolate when modeling, and discontinuity can be handled with a short length scale \cite{duvenaud2014automatic}. According to the optimized parameters in these two kernels, our model function is neither smooth nor continuous, which is not surprising as the high-dimensional dataset was taken from a highly-nonlinear physical process.


In summary, we demonstrate several applications of machine learning in relativistic laser-plasma experiments beyond optimization purposes. We have built four supervised learning regression models to predict the electron beam charge using the laser wavefront change in an LWFA experiment. All four models present similar statistics in the evaluation matrix, although Random Forest performs slightly better and Gaussian Process performs slightly worse. To justify the data quality affected by the irreproducibility in experiments, we characterize the model prediction on every single data point. Three of the models show similar performance, providing a potential way of recognizing outliers without repeated measurements. The Deep Neural Network is easy to overfit our dataset and thus not the best candidate for analyzing data quality. We include virtual measurement errors to the measured electron beam charges and DJINN is found to be the most sensitive to measurement fluctuations. Having virtual measurement errors is beneficial as it not only represents experimental conditions better but also decreases overfitting. All four models give similar feature importance rankings, which are also in line with the statistical correlation of the dataset. The Deep Neural Network requires the most computational cost, followed by DJINN, while Gaussian Process and Random Forest consume the least. Therefore, Random Forest is recommended when working with datasets from similar laser-plasma experiments.

\begin{acknowledgments}
J. Lin is supported by Air Force Office of Scientific Research (AFSOR) FA9550-16-1-0121. A. Hero is supported by the Consortium for Enabling Technologies and Innovation under Dept of Energy grant DE-NA0003921. The authors would like to thank Tony Zhang, Gang Yang, and Yirui Liu for their contribution in the preparation stage. The authors would like to thank Andrew Maris for the insightful discussion. 
\end{acknowledgments}

\nocite{*}
\bibliography{ref}

\end{document}